\documentclass[fleqn,10pt]{wlscirep}
\usepackage[utf8]{inputenc}
\usepackage[T1]{fontenc}
\title{A Network Percolation-based Contagion Model of Flood Propagation and Recession in Urban Road Networks}

\author[1,*]{Chao Fan}
\author[2]{Xiangqi Jiang}
\author[1,*]{Ali Mostafavi}
\affil[1]{Department of Civil and Environmental Engineering, Texas A\&M University, College Station, TX 77843, U.S.}
\affil[2]{Department of Computer Science and Engineering, Texas A\&M University, College Station, TX 77843, U.S.}

\affil[*]{chfan@tamu.edu, amostafavi@civil.tamu.edu}

\begin{abstract}
In this study, we propose a contagion model as a simple and powerful mathematical approach for predicting the spatial spread and temporal evolution of the onset and recession of flood waters in urban road networks. A network of urban roads resilient to flooding events is essential for provision of public services and for emergency response. The spread of floodwaters in urban networks is a complex spatial-temporal phenomenon. This study presents a mathematical contagion model to describe the spatial-temporal spread and recession process of flood waters in urban road networks. The evolution of floods within networks can be captured based on three macroscopic characteristics—flood propagation rate ($\beta$), flood incubation rate ($\alpha$), and recovery rate ($\mu$)—in a system of ordinary differential equations analogous to the Susceptible-Exposed-Infected-Recovered (SEIR) model. We integrated the flood contagion model with the network percolation process in which the probability of flooding of a road segment depends on the degree to which the nearby road segments are flooded. The application of the proposed model was verified using high-resolution historical data of road flooding in Harris County during Hurricane Harvey in 2017. The results show that the model can monitor and predict the fraction of flooded roads over time. Additionally, the proposed model can achieve 90\% precision and recall for the spatial spread of the flooded roads at the majority of tested time intervals. The findings suggest that the proposed mathematical contagion model offers great potential to support emergency managers, public officials, citizens, first responders, and other decision makers for flood forecast in road networks. 
\end{abstract}
\begin{document}

\flushbottom
\maketitle
%
%

\section*{Introduction}
Given the essential role transportation plays in emergency response, provision of essential services, and maintenance of economic well-being\cite{Ganin2017}, the resilience of urban road networks to natural hazards, especially flooding events, has received increasing attention\cite{Paprotny2018,Koks2019}. Floodwaters in urban networks propagate over time and space, inducing a great deal of spatial-temporal uncertainty vis-a-vis protective actions, such as evacuation, and rapid emergency response\cite{Wang2019}. Developing effective prediction tools to forecast the characteristics of flooding events is critical to the enhancement of urban road network resilience\cite{Fan2020}.

Multiple studies have explored the spatial-temporal properties of floods in urban networks, including impact evaluation of environmental stress\cite{Lhomme2013,Pulcinella2019,Serre2018} and cascading effects in road networks\cite{Lu2018,Guan2018}. In particular, empirical studies adopting remote sensors\cite{Mousa2016}, hydraulic data\cite{Ramsey2011}, or satellite images\cite{Dixon2006} have attempted to capture the properties of urban flooding. Temporal evolution of flood status is driven by the time-dependent profile of environmental stress, such as the duration of rainfall in hurricanes\cite{Ramsey2011}. This temporal information facilitates identification of the outbreak and inflection points for flooding in affected networks. Flooding also exhibits high spatial correlation\cite{Youssef2015} in which the co-located road segments are more likely to be flooded in the immediately succeeding time increments, indexed to digital timestamps on the model\cite{Douglas2000}. Specifically, a hydrologic study has shown that proximity to flooded areas is a significant predictor of regional flood frequency based on regionalized flood quantiles for 575 Austrian catchments\cite{Merz2005}. While empirical studies illustrated the complex spatial-temporal dynamics of floods, their capabilities for flood prediction often rely on various types of hydro-geomorphological monitoring datasets and intensive computation\cite{Nayak2005}. Due to delay and computational cost issues, the existing physics-based hydrodynamic models may not be conducive to providing timely and reliable predictions for the spatial-temporal spread of floods and the failure of road segments within short time periods\cite{Hossain2007}. 

To overcome the limitations in empirical models, machine learning techniques have been proposed and tested for predicting the spread of floodwaters in urban areas\cite{Mosavi2018,Fan2020a}. Compared to the hydrodynamic methods, machine learning models, such as multiple linear regressions\cite{Tsakiri2018}, deep neural networks\cite{Sankaranarayanan2019}, and Bayesian forecasting models\cite{Dong2019} require fewer input parameters, so that the models can be easily trained on historical flood event data. For example, Khosravi et al. tested four decision tree-based machine learning models—logistic model trees, reduced error pruning trees, naïve Bayes trees, and alternating decision trees—for flood mapping\cite{Khosravi2018}. The results show that, with adequate training, these models can achieve greater than 80\% accuracy for predicting the flooded locations. Youssef et al. integrated frequency ratio and logistic regression models to evaluate the correlation between flood occurrence and various potential factors and developed a model providing an acceptable prediction accuracy\cite{Youssef2015}. Although existing machine learning models can achieve a good predictive performance to capture the flood propagation in urban networks, these models are limited due to their dependence on large sets of historical data for model training. In addition, existing machine learning models are designed to capture only the propagation of flood in urban areas. The flood recession process, which is also important for assessing the resilience of urban networks, is often ignored by the existing machine learning models. 

Recognizing the limitations of existing models, there is a real need for mathematical models that can capture the spatial-temporal evolution of flooding without relying on a variety of input parameters and historical data such as the volume of waters and the width of the roads. Recent studies have demonstrated a surprisingly significant similarity among spreading processes in different systems, including the spread of traffic congestion in transportation, the contagion of infectious disease in populations, the diffusion of ideas in social networks\cite{Fan2020b}, as well as the evolution of flooding in urban road networks\cite{saberi2019simple,Barab2013}. Motivated by these studies, our goal in this research was to describe the floodwater spreading process using a generalized mathematical contagion models, such as classical epidemic models\cite{McCluskey2010}. Existing epidemic models offer an analytical and numerical framework to quantify and forecast multiple spread phenomena in a variety of contexts. In particular, the popular susceptible-infectious-recovered (SIR) model created the basic building blocks of epidemic modeling using infectious and recovery rates. These mathematical models have two fundamental hypotheses: compartmentalization, in which each entity is associated with a state or compartment; and homogenous mixing, in which each entity has the same chance of contacting with an inflected entity\cite{barabasi2016network}. While these hypotheses simplify the modeling of contagion by eliminating the need to know the structure of the networks, the mathematical models can still capture the temporal evolution of the fraction of infected entities in the networks very well.

Flood risk prediction is a task that should take into account both the temporal and spatial natures of the floodwater in road networks. Urban flood risk characterization requires not only knowledge of the fraction of flooded roads at each timestamp, but also needs to identify the geographic locations of flooded roads as flooding unfolds. Hence, pure mathematical models are not able to satisfy these requirements. To this end, the network percolation process has gained attention recently because it enables capture of the propagation process through the topological connectivity in networks\cite{Gao2012}. As defined in percolation theory, the spread of infection relies on the probability of the infected neighbors, in which the heterogenous mixing assumption is hold in local components of the networks\cite{Ball1997}. Specifically, the infection spreads from an initial node along edges of the percolated network\cite{Miller2009}. Hence, the percolation process reflects the “amplification” effect of neighbors and weakens global network interactions. An infection is more likely to be transmitted to those the node comes encounters. This characteristic is essential for flood propagation prediction in urban networks since it considers the spatial co-location and constraints of urban networks. Without the temporal information about the fraction of flooded roads, however, the percolation process would fail to capture the temporal evolution of flooding in urban networks.

This study proposed and tested a network percolation-based contagion model that integrates the mathematical framework and network percolation process to predict the spatial propagation and temporal evolution of flooding in urban road networks. The mathematical framework fits the temporal dynamics of the flood situations, and the percolation process identifies locations of flooded road segments. To illustrate the performance of the model, we applied it to a case study of flood evolution in Houston road networks during Hurricane Harvey. Potential control strategies are also identified based on the outcomes of the model in the case study. 

\section*{The network percolation-based contagion model}
The proposed model is composed of four components: road network modeling, flood spread characterization, flood percolation process, and model evaluation. This epidemic-like model of flood spread process in urban road networks considers both global dynamics of flood scales and local probability of affecting other co-located roads within a set of neighbors. 


\subsection*{Road network modeling}
Road networks contain hundreds of thousands of road segments, most of which are quite short, about 100 meters to 800 meters each. The traffic status information collected at the road segment level provides sufficient spatial resolution to precisely estimate the scale of flooding in traffic networks. 

\begin{itemize}
\item \textbf{Definition 1}. A \textbf{road segment} is a basic unit with a starting point and an ending point, which can be assembled into a whole road in the order of the points. 
\end{itemize}
Each point is associated with a longitude and a latitude so that it can be located to a geographical map. Then, a road segment can be represented as ($lat_s$, $lng_s$, $lat_e$, $lng_e$), where $lat_s$ and $lng_s$ are the latitude and longitude of the starting point, while $lat_e$ and $lng_e$ are the latitude and longitude of the ending point. 

Although road segments enable good resolution for understanding the situation in urban networks, flooded areas are usually not restricted in a banded road segment. Floodwater tend to start from a point and spread in all directions. In addition, massive segments and their complex connections in the networks would also cause intensive computational cost. Hence, in modeling road networks, segment-to-segment modeling is limited to follow the nature of flood spread and achieve efficient computation. To this end, grid decomposition, a commonly used method\cite{Zhou2016}, is adopted in this study to generate equal-sized grids and divide the study area into small regions (see Fig. 1).

\begin{itemize}
\item \textbf{Definition 2}. A \textbf{road grid} is a square over a rectangular projection of the geographical map.
\end{itemize}
The spatial boundary of the grid can be represented as a set of geo-coordinates ($lat_{bl}$, $lng_{bl}$, $lat_{ur}$, $lng_{ur}$), where $lat_{bl}$ and $lng_{bl}$ are the latitude and longitude of the bottom left corner, while $lat_{ur}$ and $lng_{ur}$ are the latitude and longitude of the upper right corner. To covert the road segments to grids, we apply the following criteria to the road network:
\begin{equation}
    \centering
    s_i \in grid, \;\; \textrm{if} \;\; lat_{bl} \leq \frac{lat_s+lat_e}{2} \leq lat_{ur} \quad \textrm{and} \quad lng_{bl} \leq \frac{lng_s+lng_e}{2} \leq lng_{ur}
\end{equation}
After grid decomposition, the road network assembled from road segments could be described as a series of grids. In practice, multiple factors would influence the outcome of grid decomposition. For example, a large grid would include many segments, which might lead to losing the spatial resolution. Accordingly, the model would further lose the capability of capturing the spatial spread of flood. On the other hand, a small grid that cannot cover at least one road segment will partition the network into discontinuous components, and subsequently increase the computational cost. Hence, grid decomposition requires pre-testing to ensure that the grid is able to maintain spatial resolution without unduly burdening computation. 

Once the road segments are assigned to grids, we remove grids lacking segments to reduce computational cost. The remaining grids form a network in which the grids are considered as nodes, and their shared borderline are considered as links. By doing so, we can construct an undirected grid network $G$ with average degree of $\langle k \rangle$ to represent the topology of the traffic networks. To model the flood propagation and process, we then associate a dynamical binary state variable $x$ to each of the $N$ grids (also called nodes) of the grid network $G$, such that $x_i(t) \in \{0,1\}$ represents the flood status of node $i$ at time $t$. Using a standard notation, we divide the grids into two classes, functional flow ($F$) and flooded ($C$), corresponding respectively to the values 0 and 1 of the status variable $x$. Functional flow is a state in which traffic can utilize a road (regardless of traffic level). In the context of flooding spread process, the status $C$ represents the grids which have been flooded. At each time $t$, the macroscopic flooding situation is given by the fraction of flooded grids $c(t)=\frac{1}{N} \sum_{i=1}^{N}x_i(t)$. 

\begin{figure}[ht]
\centering
\includegraphics[width=15cm]{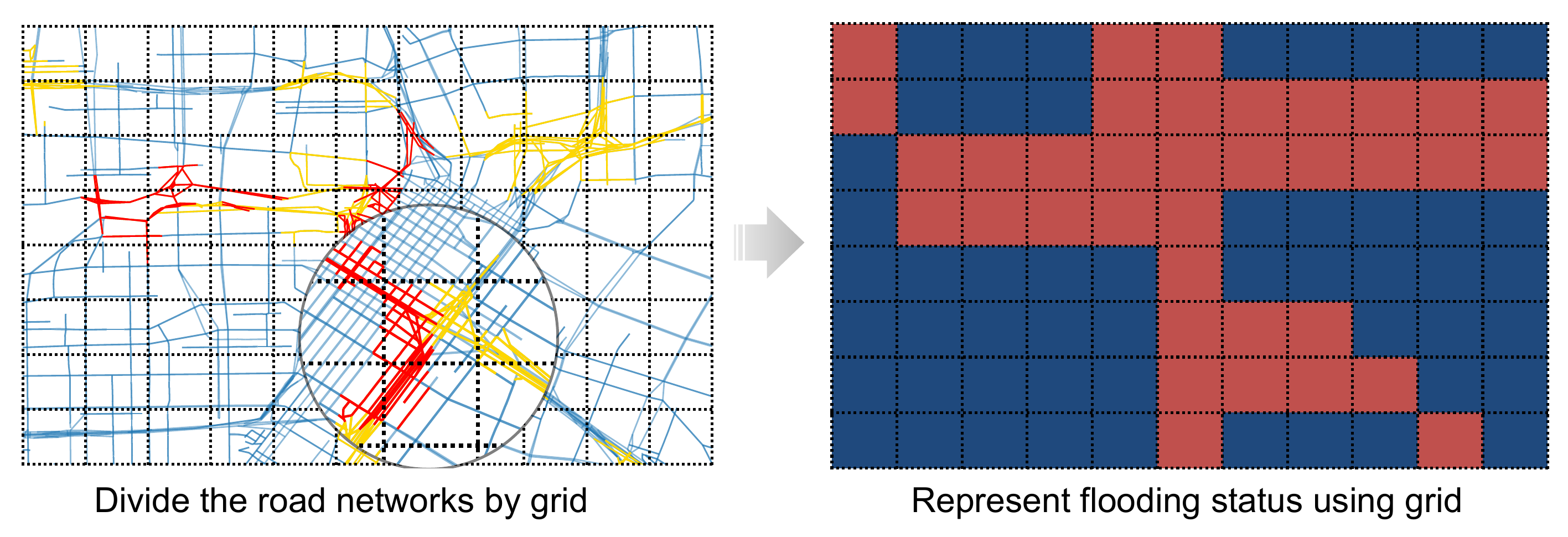}
\caption{A schema of converting flooding status from a road network to grid.}
\label{fig:Figure_1.pdf}
\end{figure}

\subsection*{Flood spread characterization}
The flood propagation and recession process are temporally and spatially variant. To capture the temporal nature of flood evolution in urban road networks, the proposed model considers macroscopic characteristics to predict the temporal evolution of floodwater spread in urban networks. 

In the first step, we define four flooding statuses for a grid: functional flow, exposed, flooded, and recovered, statuses. $C(t)$ represents the number of flooded grids in the network at time $t$; $F(t)$ represents the number of functional flow grids at time $t$; $E(t)$ represents the number of grids that are in flood incubation stage (i.e., roads in the path of approaching floodwater but on which traffic still moves) at time $t$; and $R(t)$ represents the number of grids that have recovered from flooding at time $t$. Given a grid network $G$ with average degree of  $\langle k \rangle$ and $N$ nodes, each grid in the network has functional flow at time $t=0$. That is, $F(t=0)=N$ and $C(t=0)=0$. The flood initially occurs at a set of nodes and then propagates throughout the network. From a macroscopic perspective, a grid in the undirected network is on average connected to $\langle k \rangle$ other grids. The neighbors of a flooded grid are exposed to flood at a rate of $\beta$. In modeling the temporal evolution, the connections of the grids are assumed to be homogeneous, which forms the basis to formulate a general differential equation system. Then, the probability of a flooded grid being connected to a functional-flow link is $F(t)/N$ at time $t$. Therefore, a flooded grid comes into contact with  $\langle k \rangle$ $F(t)$. Since $C(t)$ flooded grids have water flowing, each at a rate $\beta$, and the exposed grids become flooded at a fixed rate $\alpha$, the average number of new exposed grid $dE(t)$ during a unit timeframe $dt$ is determined as follows:
\begin{equation}
    \centering
    \frac{dE(t)}{dt}=\beta \langle k \rangle \frac{C(t)(N-C(t)-E(t)-R(t))}{N}-\alpha \frac{E(t)}{N}
\end{equation}
The exposed grids are not completely inundated, and traffic can still pass through, but they could be flooded within the next few timestamps. We call this stage, flood incubation stage. The defined four statuses of the road grids are the only statuses that a road grid could have. Hence, the sum of the four status variables should be $N$. That is, $N=C(t)+E(t)+F(t)+R(t)$. To capture the fraction of the grids in each status, we use the following variables: $c(t)$ to represent the fraction of flooded grids in the grid network at time $t$; $f(t)$ to represent the fraction of functional flow grids in which all road segments are accessible at time $t$; $r(t)$ to represent the fraction of recovered grids from flooding at time $t$; and $e(t)$ to represent the fraction of grids that are exposed to flooding but still in flood incubation at time $t$. Then, the differential equation for the change rate of exposed grids can be derived as follows:
\begin{equation}
    \centering
    \frac{de(t)}{dt}=\beta \langle k \rangle c(t)(1-c(t)-e(t)-r(t))-\alpha e(t)
\end{equation}
where, the product of $\beta \langle k \rangle$ is called transmission rate or transmissibility. The transmissibility can be used to measure the capability of floodwater to spread in an urban network under the same propagation rate. This also allows us to understand the effects of the topological structure of an urban network on the transmission of floodwaters. 

Since a fraction of the functional-flow grids become exposed at each timestamp, the decreasing rate of the fraction of functional-flow grids can be represented by:
\begin{equation}
    \centering
    \frac{df(t)}{dt}=-\beta \langle k \rangle c(t)(1-c(t)-e(t)-r(t))
\end{equation}
Simultaneously, in a unit timeframe, a fraction of exposed nodes would be flooded at a rate $\alpha$, and some of the flooded grids recover at a rate $\mu$. Hence, the changing rate of the fraction of flooded grids and the fraction of recovered grids can be formulated as:
\begin{equation}
    \centering
    \frac{dc(t)}{dt}=-\mu c(t)+\alpha e(t)
\end{equation}
\begin{equation}
    \centering
    \frac{dr(t)}{t}=\mu c(t)
\end{equation}
Evidently, in a large-scale network where $N$ is large, the probability of a flooded grid being connected to a functional-flow grid could be close to zero. At the macroscopic scale, the assumption of homogeneous mixing makes the prediction more tractable and robust. It should also be noted that the model does not include mortality (i.e., significantly damaged roads that would not be functional after the floodwater recedes). Since it is not often the case that a flooded road grid is severely damaged, excluding the mortality is reasonable and realistic. Based on the former constructs, the model component derived for capturing the temporal dynamics of flooding scale using macroscopic characteristics is established (Fig. 2A).

\subsection*{Network percolation process}
The spatial nature of floodwater spread in urban networks is modeled using a network percolation process. The network percolation process describes the contagion effects of a flooded grid on their network neighbors\cite{Teng2016}. Specifically, the grids whose neighbors are flooded are more likely to be flooded than the grids whose neighbors are not flooded\cite{barabasi2016network}. Similarly, floodwater is more likely to recede first from the grids whose neighbors are not flooded compared to the flooded grids whose neighbors are also flooded. To characterize this spatial aspect of flood spread, we define the probability of a node to be flooded or to be recovered in the next timestamp based on the number of flooded neighbors. The propagation and recession processes are modeled as described below, respectively (see Fig. 2B).

In the propagation process, we can obtain the number of flooded grids $N_c^{(t)}$ at a unit timestamp $t$ by the predicted fraction of flooded nodes $c(t)$ and the total number of grids in the networks. The calculation can be formulated as:
\begin{equation}
    \centering
    N_c^{(t)}=N\cdot c(t)
\end{equation}
The flooded grids at the unit timestamp t are composed of flooded grids $N_c^{(t-1)}$ at the last timestamp, plus the additional flooded grids $N_c^{(t)}$ at timestamp $t$, excluding the recovered grids. In reality, however, the fraction of recovered grids is negligible during the propagation period before reaching the flooding peak. Hence, in our model, the number of additional flooded grids, $N_p^{(t)}$, in the current timestamp $t$ is obtained by:
\begin{equation}
    \centering
    N_p^{(t)}=N_c^{(t)}-N_c^{(t-1)}
\end{equation}
As discussed earlier, the spatial pattern of the propagation process $(F \rightarrow E \rightarrow C)$ is controlled by the fraction of flooded neighbors of a grid. Hence, we assign the probability of flooding (i.e., the fraction of flooded grids among all neighbors) to the unflooded grids. Each unflooded grid will be assigned a probability $p_i^{(t)} \in \{p_1^{(t)},p_2^{(t)}, ... ,p_k^{(t)}\}$, where $k \leq N$. Then the grids are sorted based on their probabilities of flooding from high to low. The additional flooding grids at the timestamp $t$ are identified from the grids with high probabilities of flooding, subject to the number of additional flooded grids, $N_p^{(t)}$.

Like the propagation process, flood recession $(C \rightarrow R)$ can also be modeled in a way similar to the network percolation process. We first calculate the number of recovered grids $N_r^{(t)}$ at the timestamp $t$ based on the value of $r(t)$ obtained from the flood dynamics model:
\begin{equation}
    \centering
    N_r^{(t)}=N \cdot r(t)
\end{equation}
Floodwater starts receding after the peak of flooding; in actual experience, additional flooded grids usually do not occur. Hence, in the spatial prediction, the number of flooded grids $N_c^{(t)}$ at the current timestamp $t$ is equal to the number of flooded grids $N_c^{(t-1)}$ at the last timestamp $t-1$, minus the number of recovered grids $N_r^{(t)}$ at the current timestamp $t$. The calculation can be formulated as:
\begin{equation}
    \centering
    N_c^{(t)}=N_c^{(t-1)}-N_r^{(t)}
\end{equation}
In the next step, we assign the probabilities of flooding (i.e., the fraction of flooded grids among all neighbors) to the flooded grids. Each flooded grid is assigned a probability $p_i^{(t)} \in \{p_1^{(t)},p_2^{(t)},...,p_k^{(t)}\}$. That is, the probability of floodwater receding is $1-p_i^{(t)}$. In a manner different from the propagation process (in which the grids are sorted based on their probabilities of flooding from high to low), we sort the grids based on their probabilities of flooding from low to high (i.e., probabilities of floodwater receding from high to low). The recovered grids at timestamp $t$ are assigned a low probability of flooding. 

Using the above calibration in the percolation process, we can mitigate the homogeneous mixing assumptions in the flood spread characterization model by adopting local heterogenous flood probabilities and achieve a high accuracy in predicting spatial distribution of flooded grids in urban networks. 

\begin{figure}[ht]
\centering
\includegraphics[width=15cm]{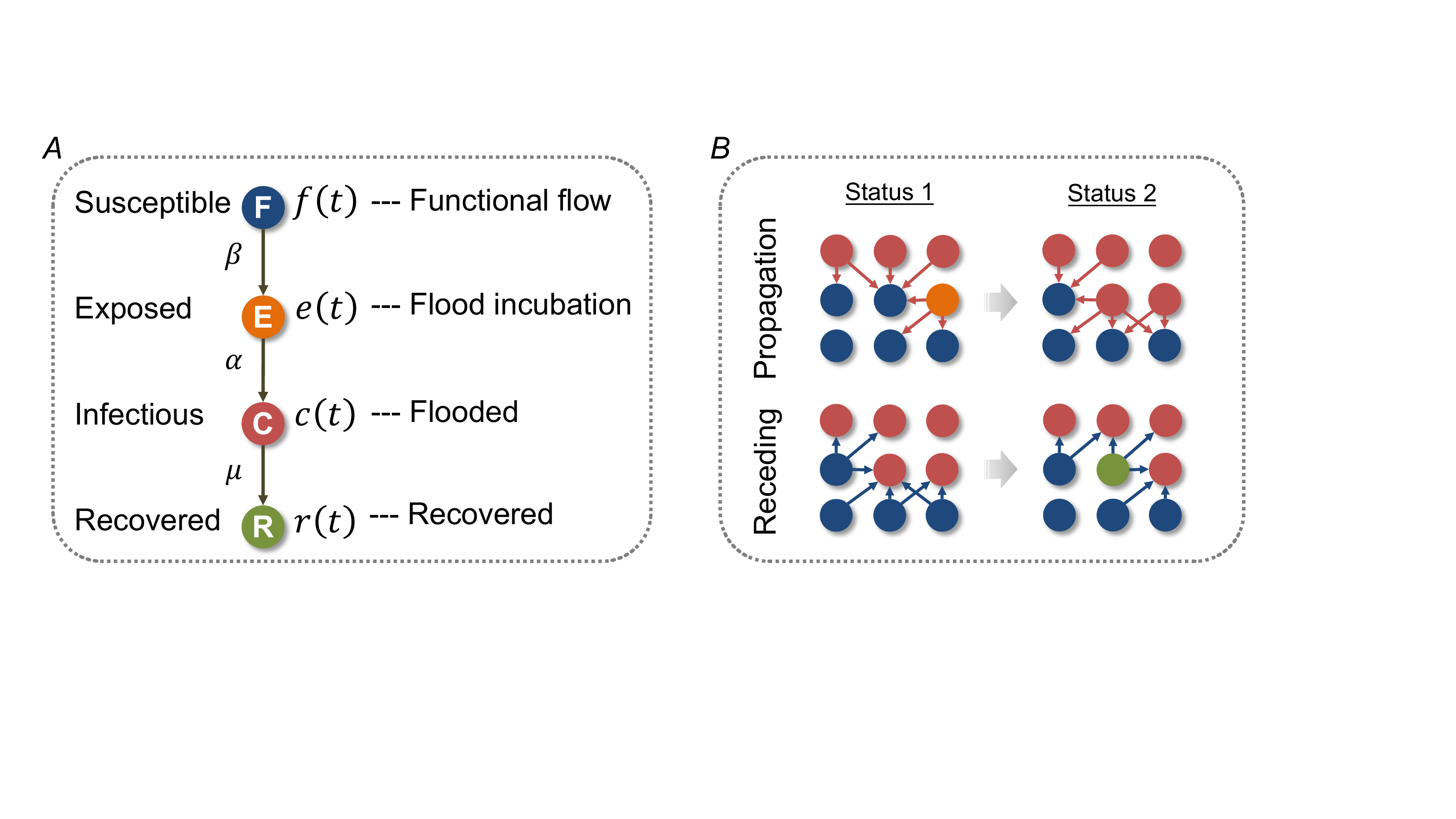}
\caption{The network percolation-based contagion model of flood propagation and recession in urban road networks. (A) The contagion model for capturing the rate of flooding grid; and (B) a schema of network percolation-based flood propagation and receding process.}
\label{fig:Figure_2.pdf}
\end{figure}

\subsection*{Model evaluation}
We employed two metrics to evaluate the performance of the model for predicting the temporal evolution and spatial propagation of flooding spread in urban networks. The first component of the model is a system of differential equations to capture the magnitude of flooded grids in networks. The objective of this component is analogous to addressing a fitting curve. Hence, we use the root mean square error (RMSE)\cite{Chai2014} to measure the error of the model: 
\begin{equation}
    \centering
    RMSE=\sqrt{\sum_{i=1}^{n}\frac{(\hat{y_i}-y_i)^2}{n}}
\end{equation}
where, $\hat{y}_i$ is the predicted value, $y_i$ is the observed value, and $n$ is the number of observations. Ignoring the division by $n$ under the square root, the formula can be considered as a formula for Euclidean distance between the vector of predicted values and observed values. Hence, the $RMSE$ is a normalized distance between the predicted outcomes and the observations which can be used to evaluate the model accuracy. This can also serve as a heuristic for a training model, which will be used in a pattern search algorithm to obtain the numerical solution for the model (see Section 3). 

The spatial nature of the flooding propagation and receding is captured by the outcomes of the flood spread characterization model and the network percolation process. The intersection between the set of predicted flooded road grids and the set of observed flooded road grids indicates the precision and recall of the model\cite{Buckland1994}. They are formulated as follows:
\begin{equation}
    \centering
    Precision=\frac{True \, positive}{True \, positive+False \, positive}
\end{equation}
\begin{equation}
    \centering
    Recall=\frac{True \, positive}{True \, positive+False \, negative}
\end{equation}
where $True \, positive$ is an outcome where the model correctly predicts the flooded class, $False \, positive$ is an outcome where the model incorrectly predicts the flooded class, and $False \, negative$ is an outcome where the model incorrectly predicts the unflooded class. The calculated precision and recall allow us to assess the performance of the model by identifying the specific correctly predicted road grids.

\section*{Results}
\subsection*{Study context and data collection}
To illustrate the application and performance of the proposed network percolation-based contagion model of flood spread, we tested the model using high-resolution data related to flooded roads in Houston during Hurricane Harvey in 2017. Hurricane Harvey was a Category 4 storm which made landfall in Houston on August 26, 2017, dissipated inland August 30, 2017\cite{Sebastian2017}. The torrential rainfall brought by Harvey caused intensive flooding in the Houston area, where the floodwaters damaged more than 290 roads and highways\cite{Ibrahim2017}. The flooding occurred August 27 through on September 4, 2017. We collected the traffic data for 19,712 road segments in Harris county from INRIX, a private company providing location-based data and analytics. The data includes the average speed for each road segment in 15-minute intervals. The timeframe of data is August 1 to September 30, covering the entire flooding period. 

In the INRIX dataset, the flooded road segments can be identified by a designation of NULL for average speed, meaning was no vehicle driving through the segment. Comparing the data before and after Hurricane Harvey, we found that the NULL average speed appears only during the flooding period. By cross-checking the flooded roads on government reports, most of the roads with NULL speed were flooded at that time. Although the average speed data was collected in 15-minute intervals, the flood situation did not evolve significantly in such a short time period. To better capture the flood propagation and receding process, we aggregated the data at 4-, 8-, and 12-hour intervals to test the performance of the model. Each interval was assigned a binary value of flooding status to the road segments based on their average speed. As documented in the model section, the value would be 0 if there is no NULL speed record, and the value would be 1 if there is a NULL speed record in the dataset (indicating functional flow status for the roads).

\subsection*{Pattern search for parameter estimation}
It is often the case that the analytical solution to the differential equation system cannot be generated. That is because the functions are usually not continuous or differentiable\cite{Davidon1991}. To efficiently estimate the parameters in the proposed model, we applied a global pattern search algorithm\cite{Raja2014} as a derivative-free numerical optimization method to fit the curves for each variable (i.e., $f(t)$, $e(t)$, $c(t)$, and $r(t)$)\cite{saberi2019simple}. The objective function in this optimization process is the $RMSE$, which will be minimized by searching for the optimal propagation rate $\beta$, recovery rate $\mu$, and exposed rate $\alpha$. To start with this algorithm, we first specified initial values for the three parameters. The change of the parameters could be either increased or decreased in each step. The algorithm would compute the $RMSE$ for each step until it finds an optimal point at which the current $RMSE$ is smaller than previous one. If the algorithm cannot find the optimal point using the current step size, it will use half of the current step size and repeat the computing process. Then all three parameters will move to the optimal values. The algorithm runs iteratively until one of the stopping criteria is met: the maximum number of iterations is reached or the step size is smaller than a certain threshold. In this study, we set the maximum number of iterations to be 2,000, and the threshold for the step size is 0.0001. 

Table \ref{tab:Table_1} shows the estimated values for the parameters and the optimized $RMSE$ for three models. The $RMSE$ increases a little bit with the increase of the length of the time intervals. That is because the fraction of road grids that are flooded (i.e., $c(t)$) is greater in longer time intervals than in shorter time intervals. But, all three models fit the flood spread patterns in road grids well since the $RMSE$ value accounts for only a very small proportion of the peak value of $c(t)$. In addition, the propagation rate $\beta$ remains the same across three models. This result indicates that the length of the time interval does not affect the capability of the model to predict flooding propagation. The flood incubation rate $\alpha$ and recovery rate $\mu$, however, decrease with the increase in time interval. That is because the number of flooded road grids increase with the increase in time intervals. Hence, a great number of road grids would be in the flood incubation state, which leads to a low rate of $\alpha$ (see Fig. 3D, E, and F). As such, the recovery rate must be small to represent the large number of flooded road grids. This result reveals that the length of the time interval has an influence on parameter estimation, but it does not significantly affect the fitting performance.
\begin{table}[ht]
\centering
\begin{tabular}{l|l l l l l}
\hline
\textbf{Length of time interval} & \textbf{$RMSE$} & \textbf{$\beta$} & \textbf{$\alpha$} & \textbf{$\mu$} & \textbf{${\beta}/{\mu}$} \\
\hline
4 hours & 0.020 & 0.9998 & 0.1267 & 0.8348 & 1.1977 \\
\hline
8 hours & 0.024 & 0.9998 & 0.0830 & 0.4073 & 2.4547 \\
\hline
12 hours & 0.027 & 0.9998 & 0.0780 & 0.2702 & 3.7002 \\
\hline
\end{tabular}
\caption{\label{tab:Table_1}Estimated values for the parameters in the contagion model.}
\end{table}

\subsection*{Prediction results}
Once the optimal values for the parameters are obtained, we can capture the temporal evolution of flood propagation and recession in urban network by showing the spread curves for each variable (i.e., $f(t)$, $e(t)$, $c(t)$, and $r(t)$) (Fig. 3D, E, and F).

Basic reproductive number $R_0={\beta}/{\mu}$ is used as a measure of the number of secondary flooded grids generated by the first flooded grid over the course of the flooding unfolding\cite{Weitz2015}. Mathematically, when $R_0$ is smaller than 1, the propagation will not occur as the recovery rate is greater than the propagation rate\cite{Chowell2004}. In the Hurricane Harvey case study, the estimated $R_0$ is greater than 1 across all three models and increases with the length of time interval (see Table \ref{tab:Table_1}). This result explains how rapidly floodwaters spread. The closer the value of $R_0$ to 1, the more stable the flooding situation is. In this case, we can observe that the flood situation would change more slightly if the time interval is shorter than 4 hours, since $R_0$ would be close to 1. When the time interval increases from 4 hours to 8 hours, the value of $R_0$ jumps from 1.20 to 2.45, meaning that in every 4- to 8-hour increment, the flood situation would change more drastically. 

In addition, the changes in $c(t)$ allow us to predict the extent of flood in the networks and critical timestamps for breakout and peak (see Fig. 3A, B, and C). The figures show the flood spread characterization in Eq.(3)–(6) are fit to the empirical flooding data very well. We can observe that at the beginning of the hurricane (the hourly timestamps in day 1 and day 2), the number of flooded grids grew exponentially between every 4-, 8-, or 12-hour increment, and reached the peak in the middle of day 3. After that timestamp on day 3, the flood receded gradually. Using the results, we can estimate the time that the urban road network needs to recover from flooding. In this case, evidently, the recovery rate was slow, which led to a long period for recovery at some severely flooded locations. 

In the next step, we examined the predictive performance for both spatial spread and temporal evolution for the flooding, shown in Figure 4. All models perform very well in terms of predicting the specific locations of flooded segments based on the flooding data at the last timestamp. The best performance of these models appears during the peak and receding period, while the performance is not particularly good at the beginning of the flooding. That is because the locations of the first initial flooded grids depend on the rainfall magnitude in different areas. After a few hours, however, the flood propagation and recession closely follow the percolation process since the majority of the initial flooded grids were identified correctly at that time. Hence, the model could predict the spatial spread of flooding after the locations of initial flooding are identified. In terms of the precision metric, the best model is the model based on a 4-hour time interval, and with the increasing of the length of the time interval, the performance of the model decreases a little bit. In terms of the recall metric, these three models show similar performance and show promise for special and temporal floodwater characteristics, especially during the peak and recession period.

The three sets of figures in Figure 5 show the true positive and false positive results for the road segments in the network. Here we converted the grids back to their segments. All segments in the flooded grids are considered flooded segments. We plotted three figures for each model at different timestamps: beginning, peak, and recession period. As we observed in the figure, the flooding initially occurred at different locations in the timestamp of the beginning period (Fig. 5, left panel). With the continuous rainfall, the floodwater propagated from the initial flooded road segments to their neighbors (middle panel of Fig. 5). After Hurricane Harvey dissipated on August 30, the flood started to recede from the road segments at the edge of the flooded area (Fig. 5, right panel). The figures show that more than 90\% of the flooded segments are captured by the proposed model, and the predictive performance decreases a bit with the increase of the length of the time interval. These findings support that the proposed model can accurately predict the spatial and temporal characteristics of flooding spread in urban networks. 

\begin{figure}[ht]
\centering
\includegraphics[width=16cm]{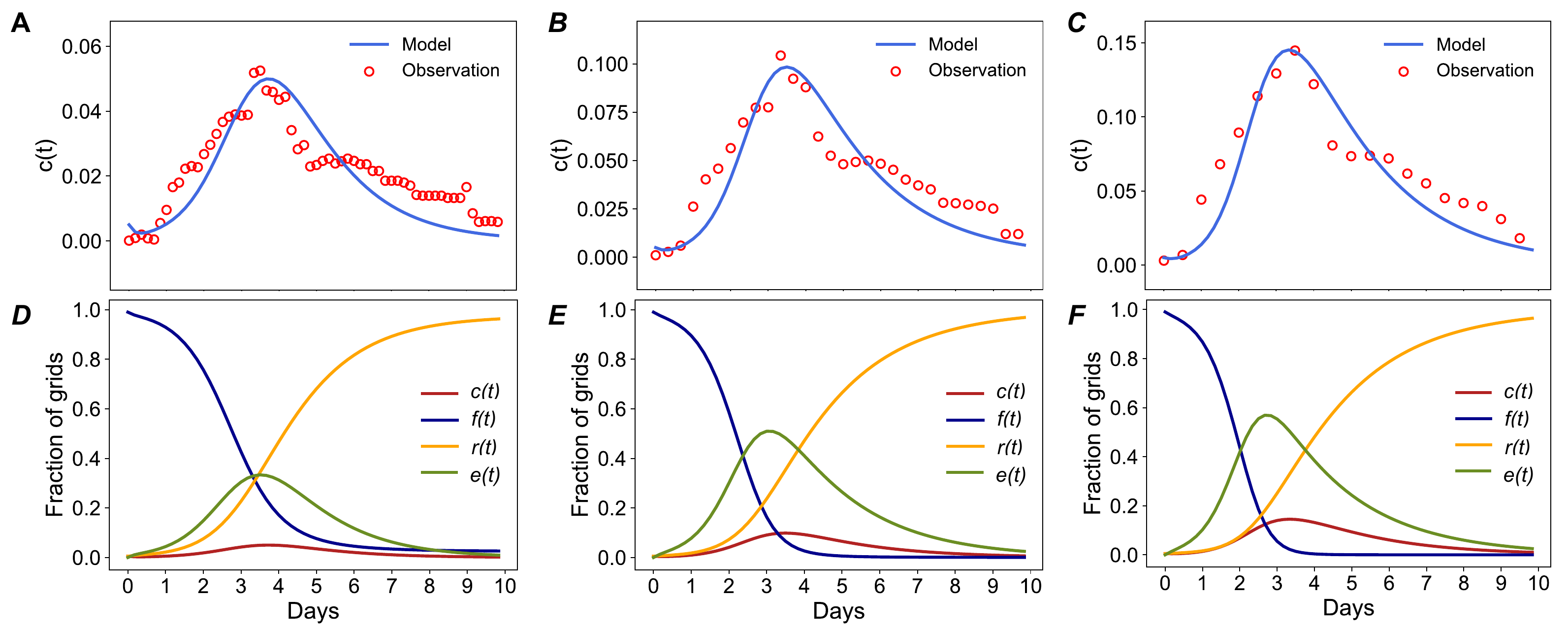}
\caption{Four-state model of flood propagation and recession in the road network, analogous to the susceptible-exposed-infected-recovered model subject to a time-varying disaster profile. Day 0 is August 26, 2017; and Day 9 is September 4, 2017.}
\label{fig:Figure_3.pdf}
\end{figure}
\begin{figure}[ht]
\centering
\includegraphics[width=15cm]{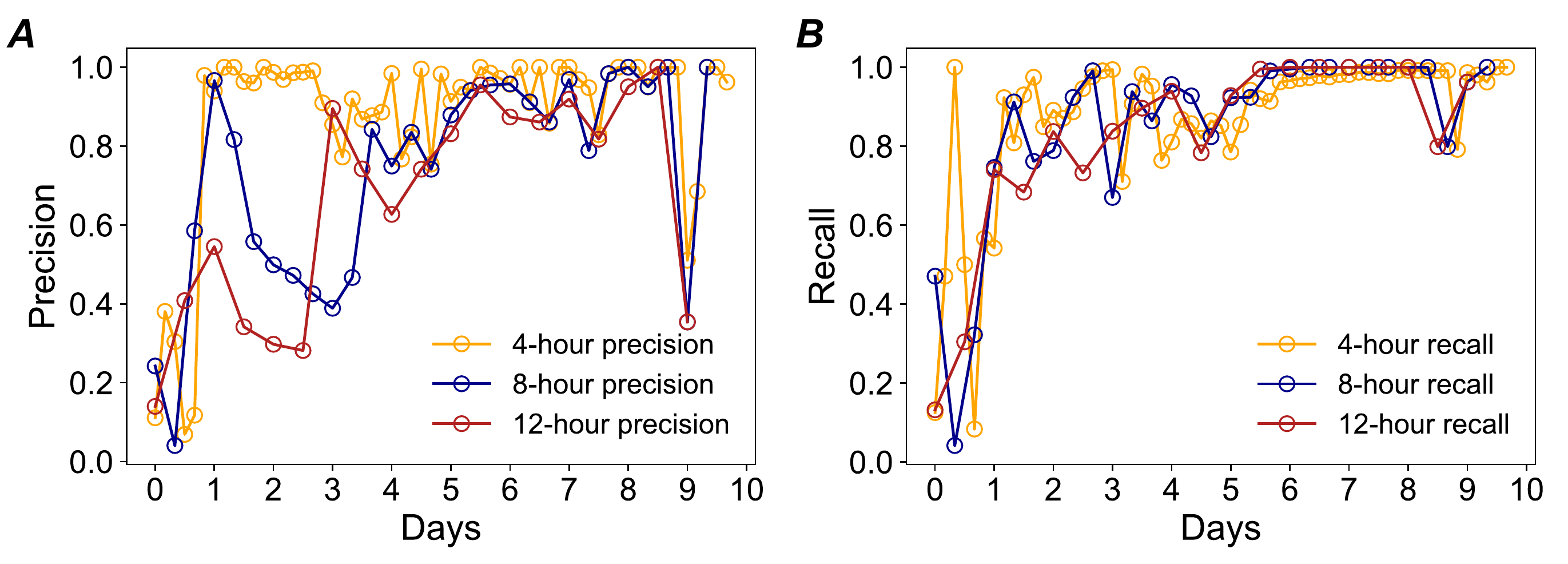}
\caption{Prediction performance of the proposed model for flood propagation and receding. (A) Precision of the model for different time intervals; and (B) Recall of the model for different time intervals. Day 0 is August 26, 2017; and Day 9 is September 4, 2017.}
\label{fig:Figure_4.pdf}
\end{figure}
\begin{figure}[ht]
\centering
\includegraphics[width=17cm]{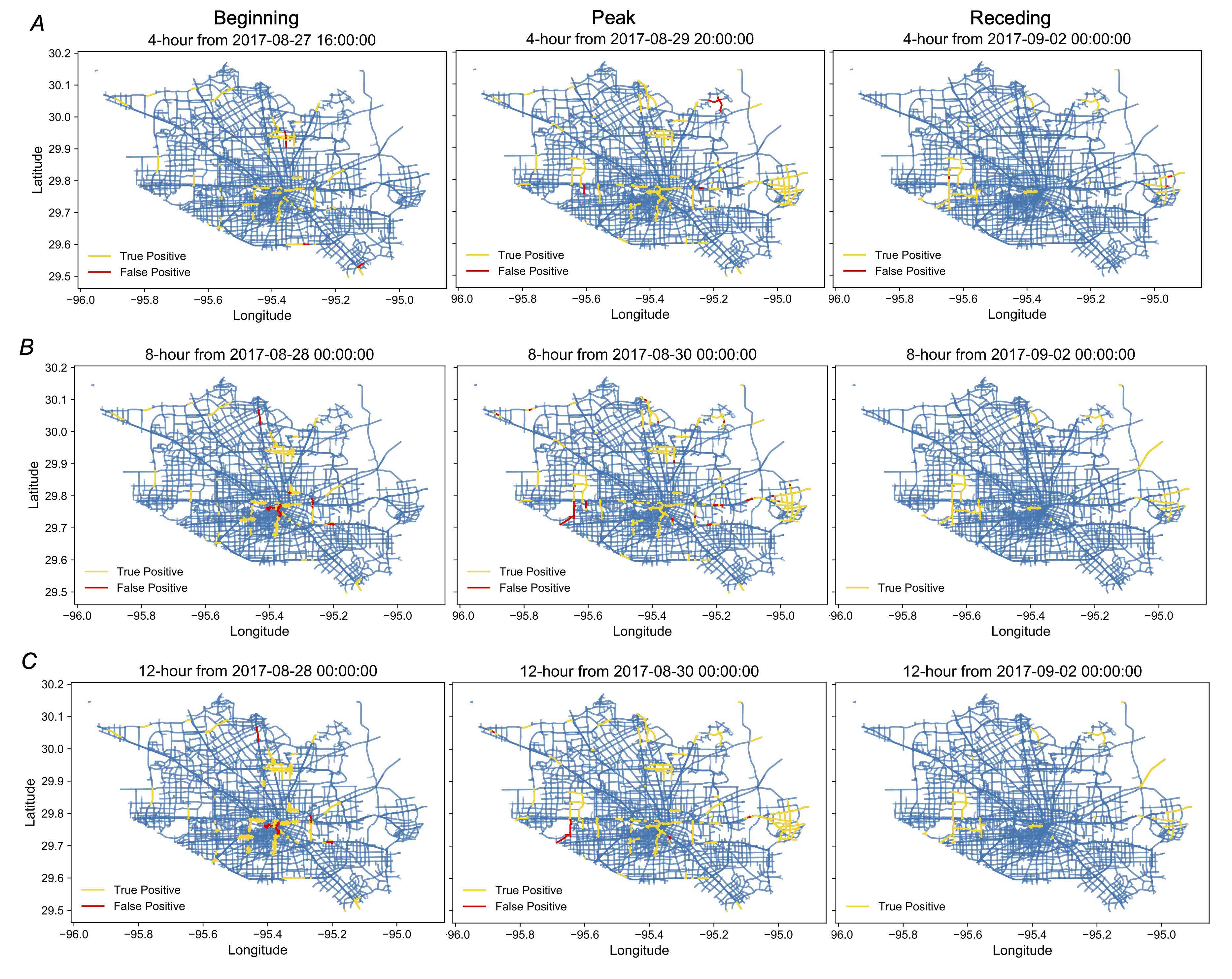}
\caption{Examples of three prediction models for different phases of the flooding period (i.e., propagating, peak, and receding phases). (A) the model for predicting the situation in next 4 hours; (B) the model for predicting the situation in next 8 hours; and (C) the model for predicting the situation in next 12 hours.}
\label{fig:Figure_5.png}
\end{figure}

\subsection*{Adaptation of the model}
The proposed network percolation-based contagion model could be used to model flood spread in other regions and flooding events. To demonstrate the adaptation of the model in other contexts, we conducted further controlled experiments to examine the adaptation of the model to different topological properties of urban networks (average degree), the propagation rate which reflects the stress of rainfall, and flood incubation rate and recovery rate. The results show the extent to which different parameters could affect the predicted outcomes. This experiment also provide evidence for supporting the generalizability of the proposed contagion model for different cities and various intensities of flooding. 

First, we tested the effect of average degree of an urban network on the flood dynamics by keeping the same values for other parameters. Figure 6A shows the changes in model results for predicting the flood status in 4-hour time intervals. With the increase of the average degree $\langle k \rangle$ of the networks, the growth rate of the flooded grids increases significantly during the propagation phase. That is because the network with a large $\langle k \rangle$ will allow the flooded road grids to infect more neighbors, which expedites the spread of flood even when the propagation rate remains the same. In addition, a large $\langle k \rangle$ would decrease the time to reach the peak of flooding since the flooded grid would have more connected neighbors. In contrast, a small $\langle k \rangle$ would lead to a longer period of flooding, although the fraction of flooded grids will not reach to the same flooding scale as the urban networks with a large $\langle k \rangle$. 

We further tested the effects of the three macroscopic parameters (i.e., $\beta$, $\alpha$, and $\mu$) on the predicted flood spread (see Fig. 6B, C, and D). On the one hand, similar to the impact of the average degree, the increase in the propagation rate $\beta$ and the flood incubation rate $\alpha$ would decrease the time to reach the peak of flooding and the number of flooded grids at the peak, but increase the time to recover to the normal situation. On the other hand, an increase in the recovery rate $\mu$ significantly decreases the number of flooded grids at the peak, but the time to reach the peak and the time to recover to normal situation are not changed significantly. The results of these experiments show that by adjusting parameters, the model can be adapted to different scenarios with various intensities of flood events and topological structures of urban networks. 
\begin{figure}[ht]
\centering
\includegraphics[width=15cm]{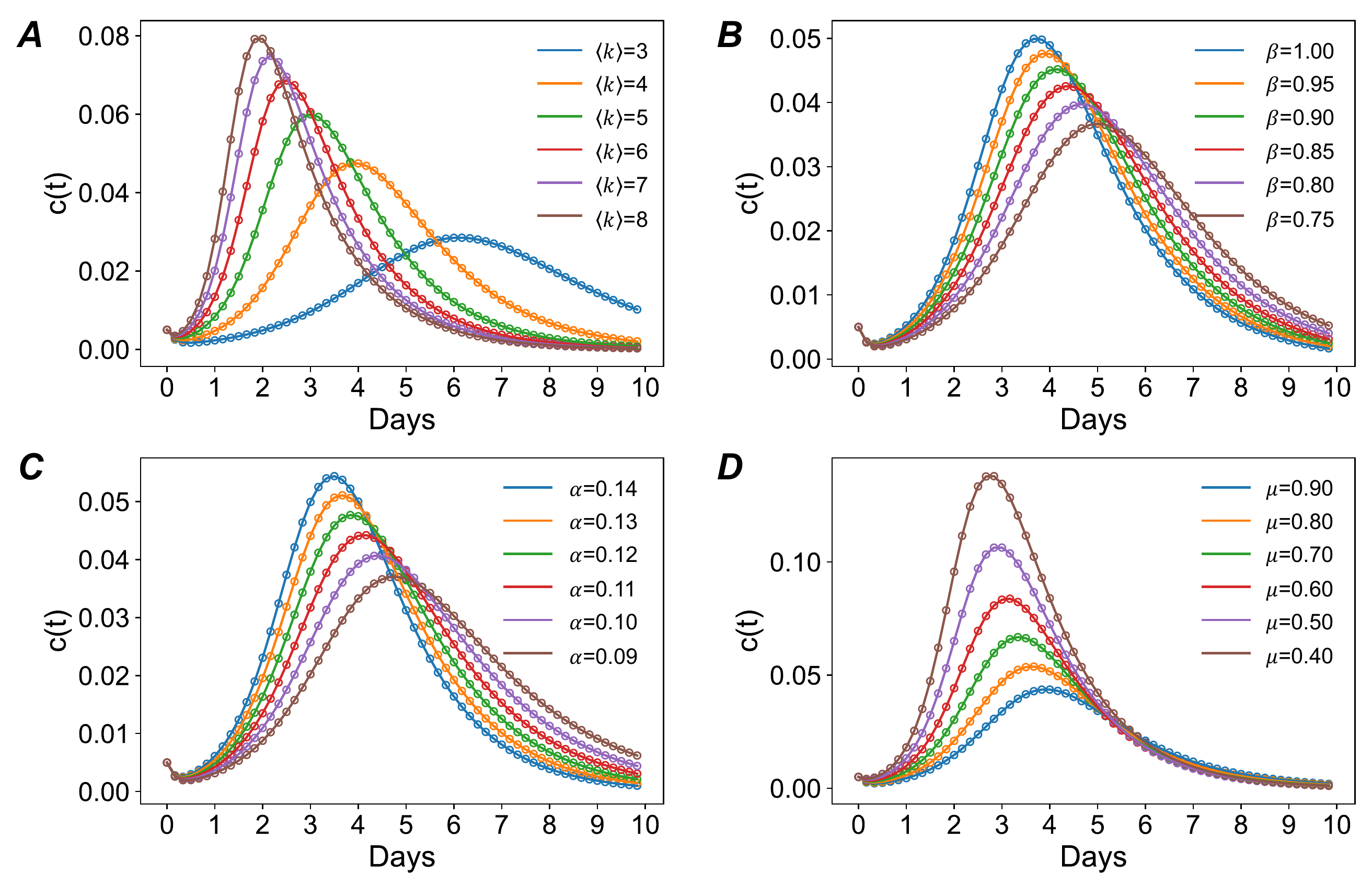}
\caption{Adaptation of the model to different topological structures of urban road networks (A); different propagation rate (B); different incubation rate (C); and different recovery rate (D), by keeping the other parameters constant.}
\label{fig:Figure_6.pdf}
\end{figure}

\section*{Discussion}
We have presented a predictive model which integrates flood spread characterization and network percolation processes to forecast the propagation and recession of flood in urban road networks. The model is formulated as a system of ordinary differential equations relying on three characteristics: flood propagation rate $\beta$, flood incubation rate $\alpha$, and recovery rate $\mu$, analogous to the SEIR model. Using the output of the model, the network percolation process is obtained to model the spatial patterns of the flooded road grids over time. The study showed the application of the proposed model in an empirical case study of urban flooding in Harris County road networks during Hurricane Harvey in 2017. The application of the model using empirical data informs about some key findings and implications for urban flood risk prediction.

First, the extent to which flooding builds up in a network and how fast it recovers are shown to be dependent on the reproductive number $R_0={\beta}/{\mu}$, which can help us infer the effective time period in response to the flooding in road networks. In the case study, we found that the reproductive number changed significantly when the time interval increased from 4 hours to 8 hours. This result indicates that flood situation evolves slightly in time intervals shorter than 4 hours, but, the situation changes dramatically for time intervals that is longer than 4 hours. Hence, to better monitor the flood in road networks, the status of roads should be observed in every 4 hours. While urban networks in different cities may have different topological properties and experience different rainfall magnitudes, the proposed model can be adapted by adjusting the parameters. Results can inform the decision makers about the spatial propagation and temporal evolution of flooding.

Second, the spatial mechanisms of flood propagation and receding in urban networks is captured based on a network percolation process. This finding highlights the contagion effects of a flooded grid on its neighbors. Since the road grids are spatial networks, different from social contact networks, the spread of flood would be limited by geographical constraints. Hence, the local contagion from a grid to another grid through the link connecting them would be the main spreading pattern. Practically, this finding provides an important implication for flood control in urban systems. One effective control strategy would be to increase drainage capacity and to build retention ponds in around roads with high likelihood of flooding. This strategy can reduce the proportion of flooded grids among the unflooded grid neighbors, which can contribute to reducing the probability of flooding in the next timestamp. 

Third, the model and its application show good performance in predicting the scale and locations of flooded road grids in disasters. In addition to the response strategies, this model could also support proactive strategies for coping with future flood events. In particular, the proposed model can be incorporated in an early warning system. The system can help officials and the public be aware of the flood situations in the coming hours so that they can perceive flooding risks surrounding them and make proactive preparations. For example, cars and buses drove through floodwater during Hurricane Harvey\cite{Bajaj}. With the predicted outcomes of this contagion model, people can be informed about roads at high risk of flooding in the next few hours. This predictive information could significantly contribute to reducing the economic loss of transportation agencies, and possibly loss of life of residents. 

Finally, the model we introduced in this paper is simple and robust and can be adapted to various phenomena. Unlike hydrodynamic models and machine learning models that rely on significant data and computational resources, the proposed mathematical model provides a simple but powerful tool for predicting the spatial-temporal evolution of flooding in urban networks. Also, the proposed network percolation-based contagion model can be adapted for modeling other network spread phenomena. Future studies can further investigate the proposed model in other general predictive tasks, such as the spread of traffic congestions in road networks\cite{Gehlot2020}, infectious diseases in human contact networks\cite{Newman2002}, and innovations in global communities\cite{FAN2020101498}. This model also has some limitations; for instance, for initial flooded segments that without flooded neighbors, it is usually difficult to predict these segments without other information. Future studies can focus on improving the model to precisely predict initial flooded road segments based on rainfall magnitude and capacity of urban drainage systems. 

\bibliography{SEIR_paper}



\section*{Acknowledgements}

This material is based in part upon work supported by the National Science Foundation under Grant Number CMMI-1846069 (CAREER), CMMI-1832662 (CRISP 2.0 Type 2), the National Academies’ Gulf Research Program Early-Career Research Fellowship, the Amazon Web Services (AWS) Machine Learning Award. The authors also would like to acknowledge the data shared by INRIX. Any opinions, findings, and conclusions or recommendations expressed in this material are those of the authors and do not necessarily reflect the views of the National Science Foundation and Amazon Web Services.

\section*{Author contributions statement}

C.F., X.J., and A.M. designed the study. X.J. and C.F. implemented the method and empirical case study. A.M. support with data acquisition. C.F. and A.M. wrote the main manuscript. All authors reviewed the manuscript.

\section*{Competing interests}
The authors declare that they have no competing interests.

\end{document}